\documentstyle[prl,aps,preprint]{revtex}
\begin{document}
\draft

\topmargin=-1.0cm

\title{Ionization of hydrogen and hydrogenic ions by antiprotons}

\author{D.R. Schultz, P.S. Krsti\'c, and C.O. Reinhold}
\address{Physics Division, Oak Ridge National Laboratory, \\
Oak Ridge, TN 37831-6373} 
\author{J.C. Wells}
\address{Institute for Theoretical Atomic and Molecular Physics, \\
Harvard-Smithsonian Center for Astrophysics, Cambridge, MA 02138} 

\maketitle

\begin{abstract}
Presented here is a description of the ionization of hydrogen and
hydrogenic ions by antiproton-impact, based on very large scale
numerical solutions of the time-dependent Schr\"odinger equation
in three spatial dimensions and on analysis of the topology of the 
electronic eigenenergy surfaces in the plane of complex internuclear
distance.  Comparison is made with other theories and very recent 
measurements.
\end{abstract}

\vspace{0.5cm}


\narrowtext

Intense, well collimated, monoenergetic beams of low energy antimatter
projectiles, such as positrons and antiprotons, have only recently 
been available for use in the study of ion-atom and ion-solid
interactions (see e.g. \cite{Schu91}). In addition to their intrinsic 
novelty, their utility stems from the fact that they allow one to probe 
the change in collision dynamics and reaction probabilities when only 
a single characteristic of the projectile is changed.  That is, 
comparison of electron- and proton-impact collisions reflects a change 
in projectile charge sign and projectile mass simultaneously, whereas 
comparison of antiproton- and proton-impact isolates the differences 
arising from the varying charge sign.  Considerable insight has been 
obtained in the last decade by comparing the ratio of double to single 
ionization \cite{Ande87}, the spectrum of electrons ejected in ionization 
\cite{Olso88}, and the variation of stopping power \cite{Agne95,Schi95} 
(the well known Barkas effect), presented by antimatter-impact. 

Prompted in large part by very recent experiments which have provided
the first measurements of the single ionization cross section by
antiproton-impact \cite{Hvel95,Knud95}, presented here is a detailed 
description of ionization in antiproton-impact of atomic hydrogen
and hydrogenic ions.  The results obtained reveal a remarkable 
difference in the mechanisms and behavior of ionization at low collision
energy.  The physical picture developed is based upon solutions of the 
time-dependent Schr\"odinger equation (TDSE) on very large, three 
dimensional, numerical lattices, and on an analysis of the quasi-molecular 
electronic eigenenergy surfaces in the plane of complex internuclear distance.
These methods circumvent uncertainties introduced by treating these 
colliding systems classically or through perturbation theory. 

Antiproton collisions with atomic hydrogen and hydrogenic ions provide 
a unique and fundamental testing ground for the development of 
non-perturbative, quantal scattering techniques.  Unlike proton-impact 
where the final state is a superposition of elastic scattering, excitation, 
ionization, and charge transfer, for antiproton-impact only the first 
three of these can be obtained.  This essential difference is obviously 
a result of the change in the charge sign of the projectile, the negatively 
charged antiproton not supporting any bound electronic states.  

In particular, the ionization cross section for proton-impact of atomic 
hydrogen displays a peak at an impact energy of about 50 keV and decreases 
below this energy due to ({\em {i}}) the lack of a strong coupling
between the relevant quasi-molecular electronic eigenstates and
the continuum, and ({\em {ii}}) the strong coupling with the charge transfer
channel.  On the other hand, it has been predicted classically \cite{Schu89}
that for antiproton-impact the cross section should not fall off
in this low energy regime. Quantum mechanically, implications that such 
a behavior should occur are provided by the fact that all the quasi-molecular
states are promoted to the continuum for small internuclear distances
\cite{Ferm47,Kimu88}.  Very recent experiments \cite{Knud95} have
nearly reached energies sufficiently low to demonstrate (or deny)
this behavior for hydrogen targets. However, an experiment employing
helium targets by the same group and part of its interpretation based 
on perturbation theory have cast some doubt on this picture \cite{Hvel95}. 
Here we provide conclusive evidence which resolves the controversy for 
one-electron systems.

To understand the antiproton-hydrogen system in more detail, consider
the quasi-molecular eigenenergy displayed as a function of distance ($R$)
between the antiproton and the proton given in Figure 1a. The character
of these states is very similar to those in a dipole potential 
$V(\vec{r}) = -\vec d\cdot \vec r/r^3$. In the limit of small $R$, the 
electron is bound to the quasi-molecule by a dipole-like potential 
produced by the two heavy particles of charge $-Z$ and $+Z$ where the 
dipole moment is $d=ZR$ and $r$ is the electronic coordinate relative 
to the center-of-mass.  There exists a critical value $d_c$ below which 
no bound states can be supported \cite{Ferm47,Kimu88}. Therefore, the 
electron cannot be bound for interparticle separations below a critical
value $R_{FT} = d_c/Z = 0.639/Z \; a.u.$, known as the Fermi-Teller
radius \cite{Ferm47}.  At this distance the electronic eigenenergies of 
$ns$-states merge with the quasi-molecular continuum edge.  In contrast, 
the eigenenergies stay bound in the united atom limit for the proton-hydrogen 
system and for antiproton--hydrogenic-ion systems (Figure 1b).

The existence of this critical radius for the antiproton-hydrogen system 
has an immediate consequence for the behavior of the ionization cross 
section for collisions with small velocity, $v$. In general, when 
$v\rightarrow 0$, colliding systems may adiabatically deform in the 
course of collision, rendering inelastic transitions improbable. This 
system provides an exception to this rule because its eigenenergies are 
degenerate with the continuum and mutually for $R \leq R_{FT}$.  More 
precisely, $R_{FT}$ is an essential singularity of the eigenenergy surface 
and for $R \leq R_{FT}$ the system becomes unstable \cite{Ovch92}, decaying 
into the continuum. This imposes an ideal lower bound on the low-energy 
ionization cross section given by $\pi R_{FT}^2 \; a.u.$, assuming a 
straight-line approximation for the internuclear motion.  

However, in reality, when $R>R_{FT}$, significant nonlocalized transitions 
are also expected between the ground state and the continuum through the 
nonadiabatic coupling operator $\partial/\partial t$ when the binding 
energy $E(R)$ of the initial quasi-molecular state satisfies 
$E(R)/v \ll 1 \; a.u.$.  This condition is fulfilled at a radius of about 
1 $a.u.$ for the 1$s\sigma$ state (Figure 1a) for 
$v {\buildrel > \over \sim} 0.1 \; a.u.$ which provides a practical limiting 
value of the cross section. Thus the adiabatic regime for the 
antiproton-hydrogen system is limited to collision energies near the 
threshold for ionization, in contrast to the proton-hydrogen system where
the onset of this regime occurs at much higher energies.

This picture is supported by the present TDSE calculations, displayed
in Figure 2. This method is an outgrowth of earlier pioneering studies
which were restricted to smaller lattice sizes and/or fewer dimensions
\cite{Maru79}.  In the present method, the electronic wavefunction and 
the Hamiltonian operator are discretized on a large 3-dimensional spatial 
lattice of points using well known pseudo-spectral methods.  The initial 
ground state of hydrogen evolves in time under the interaction with the 
projectile which moves along a classical trajectory, computed using the 
ground potential energy curve in order to account for possible trajectory 
effects. Calculating the overlap between the time-evolved state and 
lattice eigenstates allows the determination of reaction probabilities, 
and thus the ionization cross section.  The use of this method in three 
dimensions to compute cross sections for a wide range of energies has been 
a significant computational challenge, and to the best of our knowledge, 
represents the first such application of this technique in ion-atom collisions 
making a quantitative comparison with experiment. The lack of charge
transfer for antiproton-impact simplifies the calculation since one need 
not follow states bound to the projectile.  The method is described in 
much greater detail in a subsequent publication \cite{Well95}.

At high energies (above 100 keV), the TDSE result is in good agreement
with the experimental measurments \cite{Knud95}, the atomic orbital close 
coupling treatments of Toshima \cite{Tosh93} (CC$_s$), of Matir {\em et al.} 
\cite{Mati82} (CC$_m$), and of Schiwietz \cite{Schi95} (CC$_s$), and with 
the pertubative continuum-distorted-wave--eikonal-initial-state (CDW-EIS) 
approximation.  Below this energy, the close coupling treatments CC$_t$ 
and CC$_m$ overestimate the result while CC$_s$ follows the TDSE calculation 
rather closely.  The CDW-EIS approximation clearly fails at energies below 
50 keV.  Also shown in this figure is the cross section computed with the
classical trajectory Monte Carlo (CTMC) method \cite{Schu89,Cohe87}
which gives a reasonable result over a wide range of energies.  The cross 
section stays large at low collision energy since most of the classical 
orbits become unstable when the antiproton has an impact parameter smaller 
than the mean radius of the atom, in analogy with the quantal merging of the
eigenenergies with the continuum.  For larger impact parameters, the orbits 
adiabatically adjust to the perturbation and ionization is suppressed. At 
very low energies, near the ionization threshold (27.2 eV), the polarization 
of the initial electronic cloud draws the antiproton in from larger impact 
parameters and the cross section rises.  This trend is continued at even 
lower energies as the antiproton can no longer directly ionize the atom, but
as the bound state of an antiproton and proton (protonium) is formed.
This exotic capture process dominates the production of a free electron 
for very slow collisions.  

Further, both the CC$_s$ and TDSE results indicate a rather flat cross 
section between 0.2 and 100 keV, with a value in the range of approximately 
1.4 $\pi a.u.$.  This is about a factor of three larger than the cross 
section predicted on the basis of the Fermi-Teller model, indicating that 
the excess arises due to the nonlocalized character of the transitions 
even in very slow antiproton-hydrogen collisions.  That is, in the strict 
adiabatic limit, transitions are improbable except in localized \cite{Land77}
regions where the curves become degenerate, as in the antiproton-hydrogen case 
for small enough distances.  

In contrast, when an antiproton collides with a hydrogenic ion ($Z>1$) 
the united atom charge $Z-1$ is not equal to zero and the quasi-molecular 
electronic states do not become degenerate as $R \rightarrow 0$.  In this 
case, as the collision velocity increases from zero the transitions are 
localized to points where the quasi-molecular curves cross. Since terms of 
the same symmetry which support radial transitions cannot cross, these 
points are shifted to the plane of complex $R$, and are known as hidden 
crossings.  The most elaborate treatment of the transitions induced by 
these complex crossings is the so-called hidden crossings (HC) method 
\cite{Solo81} which is exact in the limit of small velocities.  In this 
approach, the transition probability rises exponentially from zero with 
increasing collision velocity, the exponent being inversely proportional 
to $v$.

All information necessary to describe ionization in slow collisions is
contained in the topology of the hidden crossings.  These crossings appear 
as branch points of the eigenenergy surface, which we have computed for 
antiprotons colliding with various hydrogenic ions through numerical 
solution of the adiabatic Schr\"odinger equation.  In Figure 3 we display 
the most important hidden crossings which lead to ionization of the ground 
states of these ions.  They are organized in the so-called $S$-superseries, 
whose terms connect pairwise and in succession the ($n, s, \sigma$) and 
($n+1, s, \sigma$) quasi-molecular states (using the spherical quantum 
numbers of the united atom.  There is a limiting point of each superseries 
when $n \rightarrow \infty$ localized closely to all other points of the 
same superseries.  This localization is described approximately by the 
size of the symbols in Figure 3 and their small values are the cause of 
a very steep diabatic promotion to the continuum from the ground state.  
Each of the systems presented here has one such series and therefore there 
is only one, strongly localized channel for ionization of a hydrogenic ion 
in slow collisions with antiprotons.  As noted above, the exception is the 
antiproton-hydrogen system for which the whole superseries degenerates to 
the Fermi-Teller limit ($R_{FT}$) on the real axis, which is an essential 
singularity rather than a branch point.  Thus, the hidden crossing method 
is not applicable. As soon as the target nuclear charge is increased beyond 
one, the $S$-superseries of branch points emerges, as shown in Figure 3.

To demonstrate the behavior of ionization for antiproton--hydrogenic-ion
systems, we have computed the TDSE, CTMC, CDW-EIS and HC result for
$p^- + He^+$, displayed in Figure 4, which exhibits the exponential drop 
of localized transitions predicted by HC at low energies.  The TDSE 
calculations follow the HC results at low energies, CTMC at intermediate 
energies, and CDW-EIS at high energies.  Evident in this figure is the upturn
of the cross section at very low energies which is due to the bending
of the antiproton trajectory caused by the Coulomb attraction with the 
He$^+$ ion.  The onset of this upturn is seen at somewhat higher energies 
than in $p^- + H$ since the long ranged Coulomb interaction is stronger 
than the polarization interaction.  A more detailed analysis of the topology 
of the hidden crossings for antiproton-impact of hydrogenic ions is given 
a forthcoming paper \cite{Krst95}.

In brief, we note that the prediction of the HC method regarding this 
pathway to ionization has not been seen for collisions involving impact 
by positive ions due to two reasons. First, ionization through charge 
transfer channels which plays a significant role for positive-ion impact 
does not take place. Second, the $S$-promotion mechanism described here 
is not the Fano-Lichten promotion mechanism which is associated with the 
passing of the centrifugal barrier (for antiproton-impact of target 
$s$-states such a centrifugal barrier is absent).  Instead, the promotion 
takes place due to the repulsive potential barrier between the electron 
and the antiproton. Such promotion was recently reported in the context 
of a model of double ionization in antiproton-helium collisions \cite{Jane95}.

Summarizing, ionization of atomic hydrogen and of hydrogenic ions by 
antiprotons is quite different from that for impact by positively charged
particles at low energy.  For atomic hydrogen these differences are 
due to the merging of the quasi-molecular electronic eigenenergies
with the continuum, and the consequent shifting of the adiabatic
regime to extremely low energies.  For He$^+$ and other hydrogenic
ions, the levels do not merge with the continuum, and the drop of
the cross section at low energies experienced in positive particle
impact is more closely obtained.  However, the topology of the complex
eigenenergy surface governing this low energy behavior is quite
different from that in the positive particle case.  TDSE and HC
calculations have been compared here with recent experimental measurements 
and the results of other approaches yielding a detailed description of 
the ionization process. Three-body classical dynamics are shown to 
approximately describe the physics of the low energy ionization problem 
whereas CDW-EIS results grossly underestimate the magnitude of the 
ionization cross section.

In this light, recent measurements \cite{Hvel95} for $p^- + He$ which 
compared favorable at low energies with CDW-EIS, which could only be 
fortuitously correct in the low energy range, are difficult to interpret.  
In particular, the measurements dropped off rapidly at low energies even 
though recent calculations of the two-electron energy levels show that 
the ground state should be promoted very near the one-electron continuum 
(i.e. the $p^- + He^+(1s)$ molecular term) at small interparticle distances 
\cite{Schi95}. That result, and an earlier less complete calculation of 
these curves \cite{Kimu88} seem in contradiction with the experimental 
findings.  Clearly, further theoretical and experimental work is need 
to elucidate the behavior in this two-electron case.

The authors acknowledge the support of the US DOE, Offices of Fusion Energy 
and Basic Energy Sciences through Contract Number DE-AC05-84OR21400, managed 
by Lockheed Martin Energy Systems. JCW has been supported by the NSF. 
We also acknowledge helpful advice from P. Gavras, M. Pindzola, A. Salin, 
J. Burgd\"orfer, S. Ovchinnikov, G. Schiwietz, and R. Janev.

\newpage

\noindent {\large {\bf Figure Captions}}

\begin{description}
\item[ ]  Figure 1. The electronic energy as a function of the internuclear
separation for $p^- + H$ and $p^- + He^+$. 

\item[ ]  Figure 2. The total ionization cross section as a function of
collision energy for $p^- + H$. The experimental measurements \cite{Knud95} 
(squares) are compared with various theoretical approaches (CTMC -- solid 
curve, TDSE -- circles, CDW-EIS -- long dashed curve, CC$_s$ -- solid curve 
connecting open circles, CC$_t$ -- dashed curve, CC$_m$ -- dot-dash curve), 
and the Fermi-Teller limit.

\item[ ]  Figure 3. Position of the S-superseries in the complex plane of 
internuclear separation for various nuclear charges ($Z$) of the hydrogenic 
ions.

\item[ ] Figure 4.  The total ionization cross section as a function
of collision energy for $p^- + He^+$.  The theoretical results are denoted
by the solid curve (CTMC), circles (TDSE), long dashed curve (CDW-EIS),
and dashed curve (hidden crossings (HC) theory).

\end{description}


\begin{references}

\bibitem{Schu91} D.R. Schultz, R.E. Olson, and C.O. Reinhold, J. Phys.
B {\bf 24}, 521 (1991);  H. Knudsen and J.F Reading, Phys. Rep. {\bf 212}, 
107 (1992).

\bibitem{Ande87} L.H. Andersen {\em et al.} Phys. Rev. A {\bf 36}, 3612
(1987); L.H. Andersen {\em et al.}, Phys. Rev. Lett. {\bf 57}, 2147 (1986);
M. Charlton {\em et al.}, J. Phys. B {\bf 21}, L545 (1988).

\bibitem{Olso88} R.E. Olson and T.J. Gay, Phys. Rev. Lett. {\bf 61},
302 (1988).

\bibitem{Agne95} M. Agnello {\em et al.}, Phys. Rev. Lett. {\bf 74}, 
371 (1995).

\bibitem{Schi95} G. Schiwietz {\em et al.} J. Phys. B submitted; private
communication.

\bibitem{Hvel95} P. Hvelplund {\em et al.}, J. Phys. B {\bf 27}, 925 (1994).

\bibitem{Knud95} H. Knudsen {\em et al.}, Phys. Rev. Lett. {\bf 74}, 4627
(1995).

\bibitem{Schu89} D.R. Schultz, Phys. Rev. A {\bf 40}, 2330 (1989).

\bibitem{Ferm47} E. Fermi and E. Teller, Phys. Rev {\bf 72}, 399 (1947);
O.H. Crawford, Proc. Phys. Soc. London {\bf 91}, 279 (1967);
I.V. Komarov {\em et al.}, in 
{\it Spheroidal and Coulomb spheroidal functions} (Nauka, Moscow, 1976);
I. Shimamura, Phys. Rev. A {\bf 46}, 3776 (1992).

\bibitem{Kimu88} M. Kimura and M. Inokuti, Phys. Rev. A {\bf 38}, 3801
(1988).

\bibitem{Ovch92} S.Yu. Ovchinnikov and J.H. Macek, {\it AIP Conference
Proceedings} {\bf 274}, VI International Conference on the Physics of Highly
Charged Ions, p.622, New York (1992).

\bibitem{Maru79} V. Maruhn-Rezwani {\em et al.}, Phys. Rev. Lett. {\bf 43},
512 (1979); C. Bottcher, Phys. Rev. Lett. {\bf 48}, 85 (1982); K.C. Kulander
{\em et al.}, Phys. Rev. A {\bf 25}, 2968 (1982); P. Gavras {\em et al.}, 
Phys. Rev. A 52, 3868 (1995).

\bibitem{Well95} J.C. Wells {\em et al.}, Phys. Rev. A (1995) submitted.

\bibitem{Tosh93} N. Toshima, Phys. Lett. A {\bf 175}, 133 (1993).

\bibitem{Mati82} M.H. Matir {\em et al.}, J. Phys. B {\bf 15}, 1729 (1982).

\bibitem{Cohe87} J.S. Cohen, Phys. Rev. A {\bf 36}, 2024 (1987).

\bibitem{Solo81} E.A. Solov'ev, Sov. Phys. JETP {\bf 54}, 893 (1981); 
E.A. Solov'ev,  Sov. Phys. Usp. {\bf 32}, 228 (1989); T.P. Grozdanov 
and E.A. Solov'ev, Phys. Rev. A {\bf 42}, 2703 (1990).

\bibitem{Land77} L.D. Landau and E.M. Lifshitz, {\it Quantum Mechanics:
Non-Relativistic Theory}, Pergamon, Oxford (1977).

\bibitem{Krst95} P.S. Krsti\'c {\em et al.}, J. Phys. B (1995) submitted.

\bibitem{Jane95} R.K. Janev {\em et al.}, J. Phys. B {\bf 28}, L615 (1995).

\end{references}
\end{document}